\documentclass[12pt]{article}

\usepackage{theorem}
\usepackage{latexsym}
\usepackage{amsmath}
\usepackage{amssymb}
\newtheorem{theorem}{Theorem}[section]

\newtheorem{lemma}[theorem]{Lemma}

\newcommand{\ket}[1]{|#1\rangle}

\begin{document}
\begin{center}
\section*{Quantum Complexity of Parametric Integration}

Carsten Wiegand\\
Fachbereich Informatik\\
Universit\"at Kaiserslautern\\
D-67653 Kaiserslautern, Germany\\
e-mail: wiegand@informatik.uni-kl.de\\
\end{center}

\begin{center}
{\bf Abstract}
\end{center}
\begin{quotation}
We study parametric integration of functions from the class 
$C^r([0,1]^{d_1+d_2})$ to $C([0,1]^{d_1})$
in the quantum model of computation. We analyze the convergence rate
of parametric integration in this model and show that it
is always faster than the optimal deterministic rate and in some cases faster 
than the rate of optimal randomized classical algorithms.
\end{quotation}

\section{Introduction}

\label{sec:1}
Summation and integration are the most famous numerical 
problems that achieved a speedup in the quantum model of computation,
compared to the optimal convergence rates of deterministic and randomized
algorithms in the classical case.

In this paper we study the problem of parametric integration, where the
integral depends on a parameter. Therefore, the solution is now a function,
so the problem carries features of both integration and approximation.

We will consider the problem from the point of view of complexity theory
and provide an analysis for the class of $r$-times continuously differentiable
functions. For this class we determine the order of the minimal error
(up to a logarithmic gap) by deriving matching upper and lower complexity
bounds.

In Section \ref{sec:2} we present the required notions from quantum 
information-based
complexity theory, recall related previous results and formulate the main 
result. Section \ref{sec:3} is devoted to the proof of the upper bound.
In Section \ref{sec:4} we prove the lower bound, and in the final Section 
\ref{sec:5} we give some comments on the results.

\section{Preliminaries}
\label{sec:2}

In this Section we formulate the problem which is investigated. Then we
give the basic definitions of quantum information-based complexity theory,
state some useful technical results and finally formulate the main result.

\subsection{Problem formulation}
\label{sec:2.1}

Let $D_1=[0,1]^{d_1}$ and $D_2=[0,1]^{d_2}$ with $d_1, d_2 \geq 0$.
On the domain $D= D_1 \times D_2$ we define the function class $C^r(D)$ for
an integer $r\geq 1$ 
as the set of all functions $f(s,t)$, for which all partial derivatives up to 
order $r$ exist and are continuous. Let $\alpha$ be a multiindex, then the 
norm $\| . \|_r$ on $C^r(D)$ is defined by 
\[
\| f \|_r := \max_{|\alpha | \leq r} \| f^{(\alpha)} \|_{C(D)}.
\]
Let $C(D_1)$ be the space of continuous functions on $D_1$ with the supremum 
norm. We consider the solution operator 
\begin{eqnarray}
\label{eq:a}
S: C^r(D) & \rightarrow & C(D_1) \notag \\
f & \mapsto & (Sf)(s) = \int_{D_2} f(s,t)\, dt .
\end{eqnarray}
This means, we study parametric integration: Integrate the family of functions
$f(s,t)$ parametrized by $s\in D_1$ over $t\in D_2$. The limiting cases 
where either $d_1=0$ (pure integration) or $d_2=0$ (pure approximation)
are formally included because they represent classical problems of
numerical mathematics. The aim of this paper is to study the intermediate
cases where $d_1\neq 0$ and $d_2\neq 0$. 

\subsection{Quantum Setting}

We use the terminology developed by Heinrich in \cite{qu:hein:2001a},
which is a translation of information-based complexity (IBC) methods to the
quantum model of computation. In order to be as selfcontained as possible,
we summarize the quantum IBC notions needed in this paper.

First, we briefly recall the standard notation of quantum computing. Let
$H_1$ be the two-dimensional complex Hilbert space $\mathbb{C}^2$ and
\[
H_m = H_1 \otimes \cdots \otimes H_1
\]
be the Hilbertian tensor product of $m$ copies of $H_1$. We use the following 
notation,
\[
\mathbb{Z}[0,N) := \{ 0, \ldots , N-1\}
\]
for $N\in \mathbb{N}$. Let 
$\mathcal{C}_m = \{ \ket{i}: i\in \mathbb{Z}[0,2^m) \}$ be the set of unit
basis vectors of $H_m$, also called classical states or basis states, and
let $\mathcal{U}(H_m)$ denote the set of unitary operators on $H_m$. 

Let $\mathcal{F}(D,K)$ be the set of mappings $f:D\rightarrow K$.
Now we introduce the notion of a quantum query. For $F\subset \mathcal{F}$
a quantum query is given by a tuple 
\[
Q=(m,m',m'',Z,\tau, \beta),
\]
where $m,m',m'' \in \mathbb{N}, m'+m'' \leq m, Z\subseteq \mathbb{Z}[0,2^{m'})$
is a nonempty subset, and
\begin{eqnarray*}
\tau: Z & \rightarrow & D\\
\beta: K & \rightarrow & \mathbb{Z}[0,2^{m''})
\end{eqnarray*}
are arbitrary mappings. The mapping $\tau$ is the coding from basis states
of one register of the quantum computer to the domain of $f\in F$, whereas 
$\beta$ is the coding of the function values from the range of $f$ to basis 
states of a second register of the quantum computer. Such a tuple $Q$ defines
a query mapping
\begin{eqnarray*}
Q: F & \rightarrow & \mathcal{U}(H_m)\\
f & \rightarrow & Q_f
\end{eqnarray*}
by
\begin{equation}
\label{eq:e1}
Q_f \ket{i}\ket{x}\ket{y} = \left\{ \begin{array}{ll}
      \ket{i}\ket{x\oplus \beta(f(\tau(i)))}\ket{y} & \text{if } i\in Z\\
      \ket{i}\ket{x}\ket{y} & \text{otherwise},
                                    \end{array} \right.
\end{equation}
where $\ket{i}\in \mathcal{C}_{m'},\ket{x}\in \mathcal{C}_{m''},\ket{y}\in \mathcal{C}_{m-m'-m''}$
(if $m=m'+m''$, we drop the last component) and $\oplus$ means addition
modulo the respective power of 2, here modulo $2^{m''}$. The total number
of qubits needed for $Q$ is $m(Q)=m$.

Suppose we are given a mapping $S: F \rightarrow G$, where $G$ is a normed
space (in this context $S$ is a general mapping). We want to approximate
$S(f)$ for $f\in F$ with the help of a quantum computer. To do so, 
we formally define the notion of a quantum algorithm. A quantum algorithm
on $F$ with no measurement is a tuple
\[
A=(Q,(U_j)_{j=0}^n),
\]
where $Q$ is a quantum query on $F$, $n\in \mathbb{N}_0$ and 
$U_j \in \mathcal{U}(H_m) \enspace (j=0, \ldots ,n)$, with $m=m(Q)$. 
Given such an $A$ and $f\in F$ we define $A_f \in \mathcal{U}(H_m)$ by
\[
A_f = U_n Q_f U_{n-1} \ldots U_1Q_fU_0 .
\]
By $n_q(A):=n$ we denote the number of queries and by $m(A)=m=m(Q)$ the
number of qubits used by $A$. We also introduce the following notation.
Let $A_f(x,y)$ for $x,y \in \mathbb{Z}[0,2^m)$be given by
\[
A_f \ket{y} = \sum_{x\in \mathbb{Z}[0,2^m)} A_f(x,y) \ket{x} .
\]
Hence $(A_f(x,y))_{x,y}$ is the matrix of the transformation $A_f$ in the
canonical basis $\mathcal{C}_{m}$.

A quantum algorithm on $F$ with output in $G$ with $k$
measurements  is a tuple
\[
A= ((A_l)_{l=0}^{k-1},(b_l)_{l=0}^{k-1}, \varphi),
\]
where $k\in \mathbb{N}$, and $A_l \; (l=0, \ldots ,k-1)$ are quantum 
algorithms on $F$ without measurement.
We set $m_l = m(A_l)$. Then $b_0 \in \mathbb{Z}[0,2^{m_0})$ and for
$1\leq l \leq k-1$, $b_l$ is a function
\[
b_l : \prod_{i=0}^{l-1} \mathbb{Z}[0,2^{m_i}) \rightarrow \mathbb{Z}[0,2^{m_l}),
\]
and $\varphi$ is a function with values in $G$,
\[
\varphi : \prod_{l=0}^{k-1} \mathbb{Z}[0,2^{m_l}) \rightarrow G.
\]
The function $\varphi$ combines the outputs of the algorithms $A_l$ to give
a final result. The functions $b_l$ determine the starting state of the
next algorithm $A_l$ depending on the results of the previous algorithms.

We also say that $A$ is a quantum algorithm with measurement(s), or just a
quantum algorithm.

Let $\mathcal{P}_0(G)$ be the set of all probability measures on $G$ whose
support is a finite set. The output of $A$ on input $f\in F$ will be an
element $A(f) \in \mathcal{P}_0(G)$ (we use the same symbol $A$ for the
mapping $A:F\rightarrow \mathcal{P}_0(G)$).
We define $A(f)$ via a sequence of random variables $(\xi_{l,f})_{l=0}^{k-1}$
(we assume that all random variables are defined over a fixed - suitably
large - probability space $(\Omega , \Sigma, \mathbb{P})$). Let now $f\in F$
be fixed and let $\xi_{l,f}$ be such that
\[
\mathbb{P}(\xi_{0,f} = x) = |A_{0,f}(x,b_0)|^2
\]
and, for $1\leq l \leq k-1$,
\[
\mathbb{P}(\xi_{l,f} = x | \xi_{0,f} = x_0, \dots , \xi_{l-1,f} = x_{l-1})
= |A_{l,f}(x, b_l(x_0, \dots , x_{l-1}))|^2 .
\]
This defines the distribution of $(\xi_{l,f})_{l=0}^{k-1}$ uniquely. Let us
define for 
$x_0 \in \mathbb{Z}[0,2^{m_0}),\dots , x_{k-1} \in \mathbb{Z}[0,2^{m_{k-1}})$
\begin{eqnarray*}
p_{A,f}(x_0,\dots , x_{k-1}) & = & |A_{0,f}(x_0,b_0)|^2 
|A_{1,f}(x_1,b_1(x_0))|^2 \dots \\
& & \dots |A_{k-1,f}(x_{k-1},b_{k-1}(x_0, \dots , x_{k-2}))|^2 .
\end{eqnarray*}
It follows that
\[
\mathbb{P}(\xi_{0,f} = x_0, \dots , \xi_{k-1,f} = x_{k-1})
= p_{A,f}(x_0,\dots , x_{k-1}) .
\]
Finally we define the output $A$ on input $f$ as
\[
A(f) = \text{dist} (\varphi (\xi_{0,f}, \dots , \xi_{k-1,f})),
\]
the distribution of $\varphi(\xi_{0,f}, \dots , \xi_{k-1,f})$.

The number $n_q(A) := \sum_{l=0}^{k-1} n_q(A_l)$ is called the number of
queries used by $A$. This is the crucial quantity for our query complexity
analysis.

Now we define the error of a quantum algorithm $A$: Let $0< \theta <1$, and
let $\zeta$ be a random variable with distribution $A(f)$. Then 
the (probabilistic) quantum error of $A$ for $S$ on input $f$ with failure 
parameter $\theta$ is defined by
\[
    e(S,A,f,\theta ):= \inf \left \{ \varepsilon \; \left| \;
   \mathbb{P}\left(\| S(f) -\zeta \|_G > \varepsilon \right) \leq \theta
      \right. \right\} .
\]
Then we put
\[
e(S,A,F,\theta ):= \sup_{f\in F} e(S,A,f,\theta )
\]
and
\[
e_n^q(S,F,\theta ):= \inf_A  \{ e(S,A,f,\theta )\; |\; n_q(A) \leq n \} .
\]
We will consider these quantities for the fixed error probability $1/4$ and
set
\[
e(S,A,f) = e(S,A,f, 1/4), \quad e(S,A,F)=e(S,A,F,1/4) ,
\]
and we define the $n$-th
minimal query error of the problem class $F$ and the mapping $S$ by
\begin{equation}
\label{eq:e2}
    e_n^q(S,F) := e_n^q(S,F,1/4) .
\end{equation}
This means that we will analyze the error rate at given cost. There is a close
connection between $e_n^q$ and the $\varepsilon$-complexity of a problem,
which is defined by
  \[
     \text{comp}_{\varepsilon}^q(S,F)
        := \min \{m\; | \;  e_m^q(S,F) \leq \varepsilon \} .
  \]
The two quantities satisfy the following relation: For all $n\in \mathbb{N}_0$,
$\varepsilon >0$ we have
\[
e_n^q(S,F) \leq \varepsilon
 \Leftrightarrow  \text{comp}_{\varepsilon_1}^q(S,F)
:= \min \{m\; | \;  e_m^q(S,F) \leq \varepsilon_1 \}
\leq n \; \; \forall \varepsilon_1 > \varepsilon .
\]

%%%%%%%%%%%%%%%%%%%%%%%%%%%%%%%%%%%%%%%%%%%%%%%%%%%%%%%%

\subsection{Tools from Quantum Complexity}

For our analysis of  parametric integration in the quantum model we
will need some statements from quantum IBC, which are now summarized:
\begin{lemma}
\label{le:a}
Let $F\subset \mathcal{F}(D, \mathbb{R})$, $l\in \mathbb{N}_0$ and let $S_k :F \rightarrow G \; \; (k=0, \ldots ,l)$
be mappings, where $G$ is a normed space. Define $S: F\rightarrow G$ by 
$S(f) = \sum_{k=0}^l S_k(f)$. Let $ \theta_0 ,\ldots ,\theta_l \geq 0$,
$n_0, \ldots ,n_l \in \mathbb{N}_0$ and put $n=\sum_{k=0}^l n_k$. Then
\begin{equation}
\label{eq:e3}
e_n^q(S,F, \sum_{k=0}^l \theta_k ) \leq \sum_{k=0}^l e_{n_k}^q(S_k,F,\theta_k) .
\end{equation}
\end{lemma}
This is a generalization of Lemma 2 from Heinrich \cite{qu:hein:2002}, which
can be proved by the same technique by just replacing the absolute value 
with the norm on $G$.
\begin{lemma}
\label{le:b}
Let $\emptyset \neq F\subseteq \mathcal{F}(D,K)$ and 
$\emptyset \neq \tilde{F}\subseteq \mathcal{F}(\tilde{D} ,\tilde{K})$. 
Let  $\Gamma: F\rightarrow \tilde{F}$ be of the following form:
there exist $\kappa, m^* \in \mathbb{N}$ and mappings
\[
\begin{array}{ll}
\eta_j :& \tilde{D} \rightarrow D \quad (j=0,\ldots ,\kappa -1)\\
\beta :& K\rightarrow \mathbb{Z}[0,2^{m^*})\\
\rho :& \tilde{D} \times \mathbb{Z}[0,2^{m^*})^{\kappa} \rightarrow \tilde{K},
\end{array}
\]
such that for $f\in F$ and $s\in \tilde{D}$
\[
(\Gamma (f))(s) = \rho(s, \beta \circ f\circ \eta_0(s), \ldots ,\beta \circ f\circ \eta_{\kappa -1}(s) ).
\]
Given a quantum algorithm $\tilde{A}$ from $\tilde{F}$ to $G$, there is a
quantum algorithm $A$ from $F$ to $G$ with
\[
n_q(A) =2\kappa n_q(\tilde{A})
\]
and for all $f\in F$
\[
A(f) = \tilde{A}(\Gamma(f)) .
\]
Consequently, if $\tilde{S} : \tilde{F} \rightarrow G$ is any mapping and
$S= \tilde{S} \circ \Gamma$, then for each $n\in \mathbb{N}_0$
\begin{equation}
\label{eq:e4}
e_{2\kappa n}^q(S,F) \leq e_n^q(\tilde{S},\tilde{F}) .
\end{equation}
\end{lemma}
The proof of Lemma \ref{le:b} can be found in Heinrich \cite{qu:hein:2002}.

We finally state some calculation rules for the query error:
\begin{lemma}
\label{rules}
Let $S,T :F\rightarrow G$ be mappings, $n\in \mathbb{N}_0$ and $e_n^q(S,F)$ be 
finite. Then it holds:
\begin{itemize}
\item[(i)] 
\begin{equation}
\label{eq:e5}
e_q^n(T,F) \leq e_q^n(S,F) + \sup_{f\in F} \| T(f)-S(f)\| .
\end{equation}
\item[(ii)] For $\lambda \in \mathbb{R}$ it holds
\begin{equation}
\label{eq:e6}
 e_q^n(\lambda S,F) =|\lambda |e_q^n(S,F) .
\end{equation}
\item[(iii)]
If $K=\mathbb{R}$ and $S$ is a linear operator from $\mathcal{F}(D,K)$
to $G$, then for all $\lambda \in\mathbb{R}$ we have
\begin{equation}
\label{eq:e7}
e_q^n(S,\lambda F) =|\lambda |e_q^n(S,F) .
\end{equation}
\end{itemize}
\end{lemma} 
The proof of this Lemma can be found in Heinrich \cite{qu:hein:2001a}.

We now cite a method how to 
increase the success probability of a quantum algorithm. 
Let $M\in \mathbb{N}$ and 
$\psi_0: \mathbb{R}^M \rightarrow \mathbb{R}$ be the median of $M$ numbers.
For a quantum algorithm $A$ we define 
$\psi_0(A^M) := \psi_0(A, \ldots ,A)$ to be the median of the 
results of $M$ repetitions of $A$.
\begin{lemma}
\label{le:4.1}
Let $T: F \rightarrow \mathbb{R}$ be a mapping and $A$ a quantum algorithm.
Then
\begin{equation}
\label{eq:e21}
e \left( T, \psi_0(A^M), f, e^{-M/8} \right) 
\leq e \left( T,A,f,\frac{1}{4} \right).
\end{equation}
\end{lemma}
A proof of this Lemma can be found in Heinrich \cite{qu:hein:2001a}.

%%%%%%%%%%%%%%%%%%%%%%%%%%%%%%%%%%%%%%%%%%%%%%%%%%%%%%%%%%%%%
\subsection{Main Results}
\label{sec:2.3}

First we recall the known results for the special cases where either
$d_1=0$ or $d_2=0$.

We use the asymptotic notation $a_n \preccurlyeq b_n$ for sequences of
nonnegative real numbers $a_n$ and $b_n$, which means that there exist
some constant $c>0$ and some $n_0 \in \mathbb{N}$ such that $a_n \leq cb_n$
for all $n\geq n_0$. If $a_n \preccurlyeq b_n $ and $b_n \preccurlyeq a_n $
then we write $a_n \asymp b_n$. We often use the same symbol $c$ for 
possibly different constants. In particular, the needed constants may depend
on $d_1, d_2$ and $r$.

Let $\mathcal{B}(G)$ denote the unit ball of a normed space $G$, i.e.
\[
\mathcal{B}(G) := \{ g\in G \; | \; \|g\|_G \leq 1 \} .
\]
Then we have the following two theorems that are important for our analysis.
The first considers quantum integration and was proved by 
Novak \cite{qu:novak:2000}.
\begin{theorem}
\label{th:2}
Let $S^{int} : C^r(D_2) \rightarrow \mathbb{R}$ be the integration operator. Then
\begin{equation}
\label{eq:e8}
e_n^q\left(S^{int}, \mathcal{B}\left(C^r(D_2)\right)\right) \asymp n^{-r/d_2-1} .
\end{equation}
\end{theorem}
Let $A_{int}(\cdot,n)$ be a sequence of quantum algorithms which is of 
optimal order, that is
\[
e\left( S^{int}, A_{int}(\cdot,n), \mathcal{B}\left(C^r(D_2)\right)\right)
\leq c n^{-r/d_2 -1} .
\]
The second theorem is concerned with approximation.
\begin{theorem}
\label{th:approx}
Let $S^{appr}$ denote the function approximation problem, that is the embedding
operator from $C^r(D_1)$ to $C(D_1)$. Then
\begin{eqnarray}
\label{eq:e9}
e_n^{det}(S^{appr}, \mathcal{B}(C^r(D_1)) & \asymp & 
e_n^{mc}(S^{appr}, \mathcal{B}(C^r(D_1)) \notag \\
& \asymp & e_n^q(S^{appr}, \mathcal{B}(C^r(D_1)) \asymp n^{-r/d_1} .
\end{eqnarray}
\end{theorem}
Here $e_n^{det}$ and $e_n^{mc}$ are the minimal deterministic and Monte
Carlo error. A detailed definition can be found in \cite{comp:sind:99}.
The proof of the rate for the quantum case is due to Heinrich \cite{qu:hein:2002b}.

We state another important result, which is in fact the
key to the integration result mentioned above. Let
\[
L_{\infty}^N := \{ g : \{0, \ldots ,N-1 \} \rightarrow \mathbb{R} \} ,
\]
with the norm $\|g\|_{L_{\infty}^N} = \max_i |g(i)|$. Then we get the optimal
rate for quantum summation, where the upper bound is from Brassard, H{\o}yer, 
Mosca and Tapp \cite{qu:ampli:2000} and the lower bound from Nayak and Wu 
\cite{qu:nayakwu:98}, with the extension that is used for our purpose
coming from Heinrich \cite{qu:hein:2001a}:
\begin{theorem}
\label{th:1}
Let $S_N : L_{\infty}^N \rightarrow \mathbb{R}$ be given by 
$S_N g = \frac1N \sum_{i=0}^{N-1} g(i)$. Then for $n<N$ there are constants 
$c_1, c_2$ not depending on $N$ such that
\begin{equation}
\label{eq:e10}
c_1 n^{-1} \leq e_n^q\left(S_N,\mathcal{B}\left(L_{\infty}^N\right)\right)
\leq c_2 n^{-1} .
\end{equation}
\end{theorem}
Note that the convergence rate does not depend on the number of summands,
so we can choose $N$ to be large enough to satisfy any needed precision
for the approximation of an integral by such a weighted sum. We denote by
$A_{sum}(\cdot ,n,N)$ a sequence of quantum algorithms with this convergence 
rate, meaning that
\[
e\left( S_N, A_{sum}(\cdot,n, N), \mathcal{B}\left( L_{\infty}^N  
\right) \right)
\leq c_2 n^{-1} .
\]

Finally we consider the rates of parametric integration. From now on
let $F$ denote the unit ball of $C^r(D_1\times D_2)$. In order to enable
comparison, we also recall the results in the deterministic and Monte Carlo
setting. A detailed analysis for the Monte Carlo setting can be found in
Heinrich and Sindambiwe \cite{comp:sind:99}. 
\begin{theorem}
\label{th:parint:det:mc}
The minimal errors of the operator $S$ from (\ref{eq:a}) behave in the 
following way:
\begin{eqnarray}
\label{eq:e11}
& e_n^{det} (S,F) & \asymp n^{-r/(d_1+d_2)} \\[6pt]
\label{eq:e12}
& e_n^{mc}(S,F)   & \asymp \left\{ \begin{array}{ll}
             n^{-(r+d_2/2)/(d_1+d_2)}(\log n)^{1/2}, & \text{if } r > d_1/2\\
             n^{-r/d_1}(\log n)^{r/d_1}, & \text{if } r < d_1/2
                           \end{array} \right.
\end{eqnarray}
and
\[
n^{-1/2}(\log n)^{1/2} \preccurlyeq e_n^{mc}(S,F) \preccurlyeq 
n^{-1/2}(\log n)^{3/2}, \quad \quad \text{if } r=d_1/2.
\]
\end{theorem}
The following theorem settles the quantum query complexity of parametric
integration and answers the question when and how much better quantum
algorithms are (as compared to deterministic and Monte Carlo methods).
The comparison is discussed in detail in Section \ref{sec:5}.
\begin{theorem}
\label{th:main}
For $d_1\neq 0, d_2\neq 0$ and $r>0$ the operator $S$ from (\ref{eq:a}) 
satisfies
\begin{eqnarray}
\label{eq:e13}
n^{-\frac{r+d_2}{d_1+d_2}} & \preccurlyeq  e_n^q(S,F) \preccurlyeq &
n^{-\frac{r+d_2}{d_1+d_2}}(\log n)^{\frac{r+d_2}{d_1+d_2}},  \enspace \text{ if } r> d_1 \quad \\
\label{eq:e14}
n^{-r/d_1} & \preccurlyeq  e_n^q(S,F) \preccurlyeq &
n^{-r/d_1}(\log n)^{r/d_1}, \qquad \quad \text{  if } r < d_1 
\end{eqnarray} 
and
\begin{equation}
\label{eq:e14a}
   n^{-1} \preccurlyeq e_n^q(S,F) \preccurlyeq n^{-1}(\log n)^{3} ,\quad 
\text{if } r = d_1.
\end{equation}
\end{theorem}

Note that in asymptotic statements we leave the logarithm unspecified,
whereas in cases in which the basis is essential we write, e.g., $\log_2 n$
or $\ln n$ to indicate base 2 or the natural logarithm. The proof of
the above theorem consists of two parts: First we prove
an upper bound for the query error $e_n^q(S,F)$. Second, we prove a lower 
bound for $e_n^q(S,F)$ which has the same order 
as the upper bound, up to a logarithmic gap.
  
%%%%%%%%%%%%%%%%%%%%%%%%%%%%%%%%%%%%%%%%%%%%%%%%%%%%%%%%%%%%%%%%%%%

\section{Upper Bound}
\label{sec:3}
Now we assume $d_1\neq 0, d_2\neq 0$ and $r>0$.
Let $k\in \mathbb{N}_0$ be fixed, and let $\Pi_k^{(1)}$ denote the partition
of $D_1$ into cubes of sidelength $2^{-k}$ with disjoint interior. Let
\[
\Lambda_k^{(1)} := \left\{ r^{-1}2^{-k}(i_1, \ldots , i_{d_1}):
0\leq i_1, \ldots ,i_{d_1} \leq r2^k \right\} 
\]
be the equidistant mesh of sidelength $r^{-1}2^{-k}$ on $D_1$. Define
\begin{equation}
\label{eq:e14b}
n_{1,k} := |\Lambda_k^{(1)}|=(r2^k+1)^{d_1} .
\end{equation}
Let
\[
P_k^{(1)} :C^r(D_1) \rightarrow C(D_1)
\]
be the $d_1$-dimensional composite
Lagrange interpolation of degree $r$ on $\Lambda_k^{(1)}$. This means, on
each cube $Q\in \Pi_k^{(1)}$ the function $P_k^{(1)}f$ is the $d_1$-dimensional
tensor product Lagrange interpolation over the nodes $Q\cap \Lambda_k^{(1)}$.
Note that the resulting function is an element of $C(D_1)$.

For $f$ fixed $P_k^{(1)}f$ is uniquely defined by 
$\{ f(s):s \in \Lambda_k^{(1)} \}$. Therefore the operator $P_k^{(1)}$ will 
also be interpreted as defined on $L_{\infty} (\Lambda_k^{(1)})$,
the space of real valued functions on $\Lambda_k^{(1)}$, equipped with
the maximum norm.

Finally we also consider the operator $P_k^{(1)}$ as acting in the 
space
$C(D_1 \times D_2)$, meaning that we interpolate with respect to the first 
component only, leaving the other one fixed. In this case $P_k^{(1)}$ is 
defined by $(P_k^{(1)}f)(s,t):=(P_k^{(1)}f(\cdot ,t))(s)$.\\
%%%%%%%%%%%%%%%%%
Let 
\[ 
\Pi_k^{(1)}=\{ Q_{kj} \}_{j=0}^{2^{d_1k}-1},
\]
that is $D_1 = \bigcup_{j=0}^{2^{d_1k}-1} Q_{kj}$ and the $Q_{kj}$ are 
cubes of sidelength $2^{-k}$ with disjoint interior. Let $s_{kj}$ be the
point in $Q_{kj}$ with the smallest Euklidean norm. We define the restriction
operator 
$R_{kj} : \mathcal{F}(D_1, \mathbb{R}) \rightarrow \mathcal{F}(D_1,\mathbb{R})$ by 
\begin{equation}
\label{op:scaling}
(R_{kj}g)(s) = \left\{ \begin{array}{ll}
                    g \left( 2^k(s-s_{kj})\right), & \text{ if } s\in Q_{kj}\\ 
                    0 & \text{ otherwise.}
                       \end{array}
               \right.
\end{equation} 
Let $v = (r+1)^{d_1}$ and let
\[
 i = i_1(r+1)^{d_1-1} + i_2(r+1)^{d_1-2} + \ldots + i_{d_1-1} (r+1) + i_{d_1}
\]
for $i=0,\ldots ,v-1$ be the representation of $i$ in base $r+1$. 
Let now $\phi_i \enspace (i=0, \dots, v-1)$ be the tensor product 
Lagrange base polynomials of degree $r$ on $D_1$ for the grid 
$\Lambda_0^{(1)}$, meaning that $\phi_i (s) = 1$ at the point 
$s = r^{-1}(i_1, \dots, i_{d_1})$ and $\phi_i(s)=0$ for all other points in 
$\Lambda_0^{(1)}$ .

Since we want to take advantage of the fast convergence of quantum summation,
we have to find a function whose integral can be approximated by quantum 
summation. This function needs a small supremum norm to give quantum summation
its full impact.
For each level $k\geq 1$ and a fixed gridpoint 
$s\in\Lambda_k^{(1)} \setminus \Lambda_{k-1}^{(1)}$
we define the detail function $f_{k,s} \in C^r(D_2)$ as the difference function
between $f$ and its approximation $P_{k-1}^{(1)} f$, both functions considered 
for this fixed $s$.

Let us consider the structure of the detail function in dependence of $f$ and
the tensor product Lagrange base polynomials. 
For fixed $s \in \Lambda_k^{(1)} \setminus \Lambda_{k-1}^{(1)}$ there is a 
cube $Q_{k-1,j(s)} \in \Pi_{k-1}^{(1)}$ with
$s \in Q_{k-1,j(s)}$ (if there are several possibilities, choose the one with 
the smallest index $j(s)$). 
The detail function has the form
\begin{equation}
\label{eq:e15}
f_{k,s}(t) = f(s,t)-(P_{k-1}^{(1)}f)(s,t) = f(s,t)-\sum_{i=0}^{v-1} 
(R_{k-1,j(s)}\phi_i)(s) f\left( s_i, t \right),
\end{equation}
where 
\begin{equation}
\label{eq:points}
s_i = s_{k-1,j(s)}+r^{-1}2^{-(k-1)}(i_1, \dots, i_{d_1}).
\end{equation}
 
The following Lemma shows that the detail function has a bounded 
$\|.\|_r$-norm. 
\begin{lemma}
\label{le:detail}
There is a constant $c>0$ such that for any function $f\in \mathcal{B}(C^r(D))$,
any integer $k\geq 1$ and any 
$s \in \Lambda_k^{(1)}\setminus \Lambda_{k-1}^{(1)}$ we have
\begin{equation}
\label{detail:bounded}
\| f_{k,s} \|_{r} \leq c.
\end{equation}
\end{lemma}
{\bf Proof:}\\
The functions $R_{kj} \phi_i \enspace (i=0, \dots, v-1)$ are the tensor
product Lagrange base 
polynomials on $Q_{kj}$ for the grid $Q_{kj} \cap \Lambda_k^{(1)}$ and
\begin{equation}
\label{eq:scalnorm}
\| \phi_i \|_{C(D_1)} = \| R_{kj} \phi_i \|_{C(Q_{kj})}
\end{equation}
for $k\in \mathbb{N}$ and $i=0, \dots, v-1$. Since 
$\sup_i \| \phi_i \|_{C(D_1)} \leq c'$ we get
\[
\| f_{k,s} \|_{r}
\leq \|f(s,t) \|_r +\sum_{i=0}^{v-1} |(R_{k-1,j(s)}\phi_i)(s)|\; \|f(s_i, t)\|_r
\leq 1+vc' .
\]
We now choose $c=1+vc'$ and the statement follows. \hfill $\square$\\
We will need the following
\begin{lemma}
\label{interpol:bounded}
Let $k\geq 0$ be an integer. Then the operator $P_k^{(1)}$, considered as acting from 
$L_\infty^{n_{1,k}}$ to $C(D_1)$ is bounded by a constant which does not
depend on $k$.
\end{lemma}
{\bf Proof:}\\
Let $z \in L_\infty^{n_{1,k}}$ with $\| z \|_{L_\infty^{n_{1,k}}} \leq 1$.
With the notation from above we infer that for $j$ such that $t \in Q_{kj}$ 
we have
\[
\left( P_k^{(1)} z \right) (t) =\sum_{i=0}^{v-1} z(i)\cdot 
\left( R_{kj}\phi_i \right) (t),
\]
and therefore
\[
\|P_k^{(1)}z\|_{C(D_1)} \leq \sum_{i=0}^{v-1} |z(i)| \cdot 
\|R_{kj}\phi_i\|_{C(Q_{kj})} \leq v\;c' < c,
\]
where $c$ is the constant from the proof of Lemma \ref{le:detail}, which is
independent of $k$.\hfill $\square$\\[8pt]

Let us now state the parameters that are needed for the proof of the upper
bound. We use a multilevel approach developed by Heinrich (see \cite{mc:hein:1997}) 
which was also used to obtain the optimal Monte Carlo rates for parametric 
integration in \cite{comp:sind:99}.
For $x\in \mathbb{R}$ the notation $\lceil x \rceil$ means the smallest 
integer greater than or equal to, and $\lfloor x \rfloor$ the greatest one 
smaller or equal to $x$. For $n\in \mathbb{N}$ we set
\begin{equation}
\label{eq:e17}
m:= \left\lfloor \frac{1}{d_1+d_2}(\log_2 n)+1 \right\rfloor .
\end{equation} 
The starting level $\widetilde{m}$ is defined by 
\begin{equation}
\label{eq:e18}
    \widetilde{m} =\left\{ \begin{array}{ll} 
        m, & \text{if } r \geq d_1\\
        0  & \text{otherwise} ,
                       \end{array} \right. 
\end{equation}
and the final level $l$ by 
\begin{equation}
\label{eq:e19}
l:= \left\{ \begin{array}{ll}
       \lceil (1+d_2/r)m \rceil , & \text{if } r \geq d_1 \\ 
       \lceil (1+d_2/d_1)m\rceil -p & \text{otherwise, where } 
           p:= \lfloor (\log_2 m)/d_1 \rfloor .
            \end{array} \right.
\end{equation}
We use $n_{1,k}$ points for the interpolation on level $k$, and we recall that
$n_{1,k} = (r2^k+1)^{d_1}$. Let
\begin{equation}
\label{eq:e19a}
M_k :=\left\lceil 8(k+3) \ln 2 + 8\ln n_{1,k} \right\rceil ,
\end{equation}
then we define the query number for quantum summation as
\begin{equation}
\label{eq:e20}
n_{2,k}:= \left\{ \begin{array}{ll}
    \lceil 2^{d_2m-\frac12 (r+d_1)(k-m)} \rceil , & \text{if } r \geq d_1,\\
    \lceil M_k^{-1} 2^{(d_1+d_2)m-d_1k -\frac12(d_1-r)(l-k)} \rceil & \text{otherwise.}
                  \end{array} \right.
\end{equation}
The number of summands for quantum summation in level 
$k\enspace (k=\widetilde{m}, \ldots , l)$ is defined as
\begin{equation}
\label{eq:e17a}
N_k :=  2^{rk d_2}n_{2,k}^{d_2} .
\end{equation}
Let us shortly describe the main idea of the proof:
In the starting level we approximate those integrals directly, which 
correspond to parameters $s$ on the roughest grid,
with the finally needed accuracy. On the finer levels
we do the same for the detail functions. Then we
interpolate the computed approximations and add them up
to get our approximation to the solution function.

%%%%%%%%%%%%%%%%%%%%%%%%%%%%%%%%%%%%
Now we prepare the discretization of the functions that will be used on the
quantum computer. To do this, we need a mapping from our function class $F$ 
to $L_{\infty}^{N_k}$.
%%%%%%%%%%%%%%%%%%%%%%%%%%%%%%%%%%%
For $k\geq \widetilde{m}$ we choose a number $m^*$ of qubits so large that
\[
2^{m^*/2-1} \geq 1
\]
and 
\begin{equation}
\label{eq:2}
2^{-m^*/2}\leq 2^{-rk}n_{2,k}^{-1} .
\end{equation}
Then we define 
\begin{eqnarray*}
    \beta: \mathbb{R} & \rightarrow & \{ 0, \ldots , 2^{m^*}-1 \}\\
 z & \mapsto & \left\{ \begin{array}{ll}
0 & \text{ if } z< -2^{m^*/2-1}\\
\lfloor 2^{m^*/2}(z+2^{m^*/2-1})\rfloor & \text{ if } -2^{m^*/2-1} \leq z <
2^{m^*/2-1}\\
2^{m^*}-1 & \text{ if } z\geq 2^{m^*/2-1} .
\end{array} \right.
\end{eqnarray*}
Furthermore we define
  \begin{eqnarray*}
    \gamma: \{ 0,\ldots ,2^{m^*}-1\} & \rightarrow &  \mathbb{R}\\
     y & \mapsto & 2^{m^*/2} y-2^{m^*/2-1}. 
  \end{eqnarray*}
%%%%%%%%%%%%%%%%%%% k = mtilde
On the starting level $\widetilde{m}$ we only have to approximate the integral
of $f$ for fixed $s\in \Lambda_{\widetilde{m}}^{(1)}$, so in this case 
we just discretize the function
$f(s,\cdot )$, which means that we have a function 
$\eta: \{ 0, \ldots ,N_{\widetilde{m}}-1 \} \rightarrow D_1 \times D_2$ which
is defined by $\eta(j)=(s, t_j)$, where the points $t_j \in D_2$ 
are node points needed for quantum summation, they will be specified  
below. Thus for the starting level we get
\begin{equation}
\label{gamma:start}
(\Gamma_{\widetilde{m},s} f)(j) = 
\gamma \left( (\beta \circ f \circ \eta(j)) \right) .
\end{equation}
%%%%%%%%%%%%%%%%%%% k > mtilde
Let now $k> \widetilde{m}$ be fixed. As already indicated, 
we will approximate the integral of the detail function
for fixed $s \in \Lambda_k^{(1)} \setminus \Lambda_{k-1}^{(1)}$ by quantum 
summation. 
For a fixed summation number $N_k$ from (\ref{eq:e17a}) we define  
\[
\eta_i : \{ 0, \ldots ,N_k-1 \} \rightarrow D_1 \times D_2 
\enspace (i=0, \ldots , v)
\] 
by
\[
\eta_i(j) := \left( s_i, t_j \right) , 
\]
where $s_v = s$ and the points $s_i \enspace (i=0,\dots , v-1)$ are  
the points from (\ref{eq:points}). The points $t_j \in D_2$ 
are again node points needed for quantum summation. Finally we define 
\[
\rho : \{0,\ldots 2^{m^*} -1\}^{v+1} \rightarrow \mathbb{R}
\]
by
\begin{equation}
\label{eq:e16}
\rho (y_0, \ldots , y_{v}) := 
\gamma (y_v) - \sum_{i=0}^{v-1} (R_{k-1,j(s)}\phi_i)(s) \gamma (y_i) .
\end{equation}
From these mappings we get the operator 
$\Gamma_{k,s} : F \rightarrow L_{\infty}^{N_k}$ by
\[
\Gamma_{k,s} f := \rho \left( (\beta \circ f \circ \eta_i)_{i=0}^{v} \right) .
\] 
This means that
\[
(\Gamma_{k,s} f)(j) = 
\rho \left( (\beta \circ f \circ \eta_i(j))_{i=0}^{v} \right) .
\]

%%%%%%%%%%%%%%%%%%%%%%%%%%%%%%%%%%%%%%%%%%%
Now we are ready to compute the query error of $S$.
By Lemma \ref{rules} we can decompose the query error into
\begin{equation}
\label{eq:decom}
e_n^q(S,F) \leq \sup_{f\in F} \| Sf -P_l^{(1)}Sf \| +e_n^q(P_l^{(1)}S,F) .
\end{equation}
So the error splits into a deterministic and a quantum part. Classical
polynomial approximation gives for $g\in C^r(D_1)$ and $k\in \mathbb{N}_0$,
(see e.g. \cite{ciar:1976}, Chapter 3.1)
\begin{equation}
\label{eq:app}
\| g -P_k^{(1)}g \|_{C(D_1)} \leq c 2^{-rk} ,
\end{equation}
so for the deterministic part in (\ref{eq:decom}) we get
\begin{equation}
\label{eq:app2}
\| Sf -P_l^{(1)}Sf \|_{C(D_1)} \leq c 2^{-rl} .
\end{equation}
Next we consider the quantum part of (\ref{eq:decom}).
Let
\[
j= j_1 b^{d_2-1} + j_2 b^{d_2-2} + \ldots + j_{d_2-1} b +j_{d_2} 
\]
for $j=0, \ldots, N_k-1$, where $b=2^{rk}n_{2,k}$. Let the node points
for the quantum summation be defined as
\[
t_j := \left( \frac{j_1}{b} , \frac{j_2}{b}, \ldots, \frac{j_{d_2}}{b} 
\right) . 
\]
For $k>\widetilde{m}$ we define the operators $J_{k,s}: F \rightarrow \mathbb{R}$
by
\begin{equation}
\label{eq:e23}
J_{k,s} f := \frac{1}{N_k} \sum_{j=0}^{N_k-1} f_{k,s} (t_j) ,
\end{equation}
which is the rectangle rule with $N_k$ points for $f_{k,s}$.
Next we define, also for $k>\widetilde{m}$, operators 
$U_{k,s} : F \rightarrow \mathbb{R}$ by
\begin{equation}
\label{eq:e24}
U_{k,s}f = \int_{D_2} f_{k,s}(t)\;dt .
\end{equation}
Since the accuracy of the rectangle rule with $N_k$ points in dimension $d_2$
is of the order $N_k^{-1/d_2}$ for functions with bounded first 
derivatives, we get by (\ref{detail:bounded}) and (\ref{eq:e17a})  
\begin{equation}
\label{eq:1}
|U_{k,s}(f) - J_{k,s}(f) | \leq c 2^{-rk} n_{2,k}^{-1} .
\end{equation}
By definition of the discretization operator $\Gamma_{k,s}$ we get 
for $|z|\leq 1$
\[
\gamma (\beta(z)) \leq z \leq \gamma(\beta(z)) +2^{-m^*/2},
\]
and by (\ref{eq:2}) this implies that
\begin{eqnarray}
\label{eq:accuracy}
& & \left| \Gamma_{k,s}(f)(j) - f_{k,s}\left( t_j \right) \right| \notag \\
& \leq & \left| f(s,t_j) -\gamma (\beta ( f(s, t_j))) \right| +\sum_{i=0}^{v-1}
 |R_{k-1,j}\phi_i(s)| \left| f(s_i, t_j) -\gamma (\beta ( f(s_i, t_j))) \right|
\notag \\
&\leq & c 2^{-rk}n_{2,k}^{-1} .
\end{eqnarray}
From the discretization accuracy of $\Gamma_{k,s}$ and (\ref{eq:app}) we 
also infer that
\begin{eqnarray}
\label{eq:e25}
\left\| \Gamma_{k,s}(f) \right\|_{L_{\infty}^{N_k}} 
& \leq &
\left\| \left( f_{k,s}(t_j) \right) \right\|_{L_{\infty}^{N_k}}
+ \left\| \left( f_{k,s}(t_j) \right) - \Gamma_{k,s}f \right\|_{L_{\infty}^{N_k}} \notag \\
& \leq & 
c 2^{-r(k-1)} + c n_{2,k}^{-1}2^{-rk} \leq c_1 2^{-rk},
\end{eqnarray}
which implies
\begin{equation}
\label{eq:3}
\Gamma_{k,s}(F) \subseteq c_1 2^{-rk} \mathcal{B}\left( L_{\infty}^{N_k} \right).
\end{equation}
From (\ref{eq:accuracy}) it also follows that
\begin{equation}
\label{eq:4}
| S_{N_k} \Gamma_{k,s} f -J_{k,s} f| \leq
\frac{1}{N_k} \sum_{j=0}^{N_k-1} |(\Gamma_{k,s}f)(j) - f_{k,s}(t_j)| 
\leq c2^{-rk}n_{2,k}^{-1} .
\end{equation}

Now we calculate the error of the integration of the $f_{k,s}$ 
on $\Lambda_k^{(1)} \setminus \Lambda_{k-1}^{(1)}$. 
We get with Lemma \ref{rules},
(\ref{eq:4}), Theorem \ref{th:1}, Lemma \ref{le:b} and (\ref{eq:3})
\begin{eqnarray}
\label{eq:interror}
& & e_{2(v+1)n_{2,k}}^q(U_{k,s},F) \notag \\
 & \leq & \sup_{f\in F}|U_{k,s}(f) -J_{k,s}(f) | +
e_{2(v+1)n_{2,k}}^q\left(J_{k,s}, F \right)  \notag \\
 & \leq &  \sup_{f\in F}|U_{k,s}(f)-J_{k,s}(f) | 
          +\sup_{f\in F}| S_{N_k} \Gamma_{k,s} f -J_{k,s} f| 
          +e_{2(v+1)n_{2,k}}^q\left( S_{N_k} \Gamma_{k,s}, F \right) \notag \\
& \leq & c2^{-rk}n_{2,k}^{-1} + e_{n_{2,k}}^q (S_{N_k}, c_1 2^{-rk}\mathcal{B}(L_{\infty}^{N_k})) \notag \\
& \leq & c2^{-rk}n_{2,k}^{-1} +c_1 2^{-rk} e_{n_{2,k}}^q (S_{N_k}, \mathcal{B}(L_{\infty}^{N_k})) \notag \\ 
& \leq & c2^{-rk}n_{2,k}^{-1}+c 2^{-rk} n_{2,k}^{-1} \notag \\
& \leq & c2^{-rk}n_{2,k}^{-1}. 
\end{eqnarray}
%%-------------------------------------------------------------------------
With the help of this result we can now investigate the error of the operator
$P_l^{(1)}S$. Since
\begin{equation}
\label{interpolsum}
P_l^{(1)}S = P_{\widetilde{m}}^{(1)}S + \sum_{k=\widetilde{m}+1}^l
(P_k^{(1)}-P_{k-1}^{(1)})S ,
\end{equation}
we investigate the error of the operator 
\[
(P_k^{(1)}-P_{k-1}^{(1)})S: F \rightarrow C(D_1).
\] 
We define 
\begin{equation}
\label{eq:theta}
\theta_k := 2^{-(k+3)}, k=\widetilde{m} ,\ldots ,l .
\end{equation}
Then we set
\begin{equation}
\hat{n}_{\widetilde{m}} :=  M_{\widetilde{m}} n_{1,\widetilde{m}} n_{2,\widetilde{m}},
\end{equation}
and for $k = \widetilde{m} +1, \dots, l$ we set
\begin{equation}
\label{eq:nhat}
\hat{n}_k :=  M_k (n_{1,k}-n_{1,k-1})2(v+1)n_{2,k}.
\end{equation}

Let $A_{k,s}$ be a quantum algorithm that computes an approximation to
$U_{k,s}$ on $F$ with the rate from (\ref{eq:interror}) and let $\zeta_{k,s}$ 
be a random variable with distribution $A_{k,s}$. 
We define a random variable $\xi_k$ with values in $L_\infty^{n_{1,k}}$ as
follows: For $s\in \Lambda_k^{(1)}\setminus \Lambda_{k-1}^{(1)}$ we let
$\xi_k(s)$ be the median of $M_k$ independent copies of $\zeta_{k,s}$, 
that is, we repeat $A_{k,s} \enspace M_k$ times. 
For $s\in \Lambda_{k-1}^{(1)}$ we set $\xi_k(s):=0$. Since
$(U_{k,s}f)(t)=0$ for $s\in \Lambda_{k-1}^{(1)}$, by this choice 
we establish an error of zero in these points. We have
\begin{equation}
\label{eq:factorize}
(P_k^{(1)}-P_{k-1}^{(1)})(Sf) = P_k^{(1)}\left( (I-P_{k-1}^{(1)})(Sf) \right)
= P_k^{(1)}(U_{k,s})_{s\in \Lambda_{k}^{(1)}} ,
\end{equation}
where on the right hand side $P_k^{(1)}$ is considered as acting on
$L_{\infty}^{n_{1,k}}$.
This means that because we can interchange interpolation with respect to the first component and integration with respect to the second component, we indeed compute
an approximation to $(P_k^{(1)}-P_{k-1}^{(1)})(Sf)$ by
$(P_k^{(1)}-P_{k-1}^{(1)})\xi_k$.

%%%%%%%%%%%%%%%%%%%%%%%%%%%%%%%%%%%%%%%%%%%%%%%%%%%%
By Lemma \ref{le:4.1} and (\ref{eq:interror}),
\begin{equation}
  \mathbb{P}(|U_{k,s}(f) -\xi_{k} (s)| > 
     c2^{-rk}n_{2,k}^{-1} ) \leq e^{-M_k/8} .
\end{equation}
Consequently,
\begin{eqnarray}
\label{probability}
  & & \mathbb{P}(|U_{k,s}(f) -\xi_{k} (s)| \leq 
     c2^{-rk}n_{2,k}^{-1}\; \; \forall \; 
     s \in \Lambda_k^{(1)} )  \notag \\
  & \geq & 1-n_{1,k}e^{-M_k/8} \geq 1-2^{-(k+3)} = 1-\theta_k 
\end{eqnarray}
by (\ref{eq:e19a}).
%%%%%%%%%%%%%%%%%%%%%%%%%%%%%%%%%%%%%%%%%%%%%%%%%%%%
From Lemma \ref{interpol:bounded}, 
(\ref{eq:nhat}), (\ref{eq:factorize}) and (\ref{probability}) we obtain for 
the query error of the operator $(P_k^{(1)}-P_{k-1}^{(1)})S$,
\begin{equation}
\label{eq:e26}
  e_{\hat{n}_k}^q \left( (P_k^{(1)}-P_{k-1}^{(1)})S , F, \theta_k \right) 
 \leq  c 2^{-rk}n_{2,k}^{-1}.
\end{equation}
We use Lemma \ref{le:a} to calculate the error of $P_l^{(1)}S$. From 
(\ref{eq:theta}) we get $\sum_{k=\widetilde{m}}^l \theta_k \leq 1/4$, hence
with Lemma \ref{le:a}, Lemma \ref{interpol:bounded}, (\ref{interpolsum}) 
and with 
\begin{equation}
\label{eq:cost}
\tilde{n}:= \sum_{k=\widetilde{m}}^{l} \hat{n}_k ,
\end{equation}
we get
\begin{eqnarray}
\label{eq:summe}
e_{\tilde{n}}^q(P_l^{(1)}S, F, 1/4) & \leq & 
e_{\hat{n}_{\widetilde{m}}}^q((P_{\widetilde{m}}^{(1)}S,F, \theta_{\widetilde{m}}) \notag \\
& & +\sum_{k=\widetilde{m}+1}^l e_{\hat{n}_k}^q((P_k^{(1)}-P_{k-1}^{(1)})S,F, \theta_k) .
\end{eqnarray}
%%%%%%%%%%%%%%%%%%%%%%%%%%%%%%%%%%%%%%%%%%%%%%%%%%%%%%%%%%%%%%
Now we consider the different cases:\\[8pt]
{\em 1.  $r<d_1$.} For the error on the starting level $\widetilde{m}=0$ we 
can make direct use of Theorem \ref{th:1} and, in this case using the
operator from (\ref{gamma:start}), we get
\[
e_{\hat{n}_0}^q \left( P_0^{(1)}S, F, \theta_0 \right) \leq c \left( n_{2,0} \right)^{-1} 
\]
by Theorem \ref{th:1}, Lemma \ref{interpol:bounded} and a similar probability 
argument as above. Now with (\ref{eq:summe}) and (\ref{eq:e26}) we get
\begin{eqnarray*}
& &  e_{\tilde{n}}^q \left( P_l^{(1)}S, F, 1/4 \right) \\
& \leq & c \left( n_{2,0} \right)^{-1}+ 
\sum_{k=1}^l c 2^{-rk}n_{2,k}^{-1}\\
& \leq & c (\log_2 n) \sum_{k=0}^l 
2^{-(d_1+d_2)m+d_1k+\frac12(d_1-r)(l-k)-rk}\\
& \leq & c (\log_2 n) 2^{-(d_1+d_2)m} \sum_{k=0}^l
2^{\frac12(d_1-r)(l+k)} .
\end{eqnarray*}
From this we get with the help of the geometric sum formula
\[
e_{\tilde{n}}^q \left( P_l^{(1)}S, F, 1/4 \right)
\leq c (\log_2 n) 2^{-(d_1+d_2)m+(d_1-r)l}
\]
and with $2^{d_1l-d_2m} \asymp 2^{d_1m-\log_2 m}$ we arrive at
\[
e_{\tilde{n}}^q \left( P_l^{(1)}S, F, 1/4 \right)
\leq c 2^{-rl} \leq c n^{-r/d_1} (\log_2 n)^{r/d_1} .
\] 
For the deterministic part of the error we get by the choice of $l$
\[
\| Sf -P_l^{(1)}Sf \| \leq c2^{-rl} \leq c n^{-r/d_1} (\log_2 n)^{r/d_1},
\]
which by (\ref{eq:decom}) gives the desired rate for $r<d_1$.\\[8pt]
%%%%%%%%%%%%%%%%%%%%%%%%%%%%%%%%%%%%%%%%%%%%%%%%%%%%%%%%
{\em 2. $r\geq d_1$.}
To calculate the error on the starting 
level $\widetilde{m}=m$ we use Theorem \ref{th:2} and (\ref{eq:e20}) and 
again with the probability argument from above we get
\begin{equation}
\label{eq:e27}
e_{\hat{n}_m}^q ( P_m^{(1)}S, F, \theta_m ) \leq  
c \left( n_{2,m} \right)^{-r/d_2-1} \leq c n^{-(r+d_2)/(d_1+d_2)} .
\end{equation}
With (\ref{eq:summe}) and (\ref{eq:e26}) we get  
\begin{eqnarray*}
& &  e_{\tilde{n}}^q \left( P_l^{(1)}S, F, 1/4 \right) \\
& \leq & c n^{-(r+d_2)/(d_1+d_2)}+
\sum_{k=m+1}^l c 2^{-rk}n_{2,k}^{-1}\\
& \leq & c n^{-(r+d_2)/(d_1+d_2)} + c \sum_{k=m+1}^l 
2^{-rk}2^{-d_2m+\frac12 (r+d_1)(k-m)}\\
& \leq &  cn^{-(r+d_2)/(d_1+d_2)} + c2^{-(r+d_2)m} 
\sum_{k=m+1}^l 2^{-\frac12 (r-d_1)(k-m)} \\
& \leq & cn^{-(r+d_2)/(d_1+d_2)}\sum_{k=m+1}^l 2^{-\frac12 (r-d_1)(k-m)} .
\end{eqnarray*}
For $r>d_1$ the sum is bounded by a constant, and for $r=d_1$ the sum
gives an additional factor of $\log n$.
By the choice of $l$ we get
\[
\| Sf -P_l^{(1)}Sf \| \leq c2^{-rl} \leq c n^{-(r+d_2)/(d_1+d_2)},
\]
and with (\ref{eq:decom}) we arrive at
\begin{equation}
\label{r_gr_d1}
e_{\tilde{n}}^q(S,F) \leq cn^{-(r+d_2)/(d_1+d_2)} 
\end{equation}
for $r>d_1$ and
\begin{equation}
\label{r_gl_d1}
e_{\tilde{n}}^q(S,F) \leq cn^{-1} \log n 
\end{equation}
for $r=d_1$.\\

%%%%%%%%%%%%%%%%%%%%%%%%%%%%%%%%%%%%%%%%%%%%%%%%%%%%%%%%% 
Finally we estimate the number of queries $\tilde{n}$ that are needed 
to obtain the desired precision.
Since the total number of queries is
\[
\tilde{n} = M_{\widetilde{m}}n_{1,\widetilde{m}}n_{2,\widetilde{m}}
        + 2(v+1) \sum_{k=\widetilde{m}+1}^{l} M_k (n_{1,k}-n_{1,k-1})n_{2,k}, 
\]
we get for $r < d_1$
\begin{eqnarray*}
\tilde{n} & = & M_{0}n_{1,0}n_{2,0} +2(v+1)\sum_{k=1}^{l} M_k (n_{1,k}-n_{1,k-1})n_{2,k}\\ 
& \leq & c \sum_{k=0}^{l} M_k n_{1,k} n_{2,k}\\
& \leq & c \sum_{k=0}^{l} 2^{d_1k} 2^{(d_1+d_2)m -d_1k -\frac12(d_1-r)(l-k)}\\
& \leq & c 2^{(d_1+d_2)m} \sum_{k=0}^{l} 2^{-\frac12 (d_1-r)(l-k)}
\asymp n ,
\end{eqnarray*}
and for $r> d_1$ we get
\begin{eqnarray*}
\tilde{n} & = & M_{m}n_{1,m}n_{2,m}+2(v+1) \sum_{k=m+1}^{l} M_k (n_{1,k}-n_{1,k-1})n_{2,k}\\ 
& \leq & c \sum_{k=m}^{l} M_k n_{1,k} n_{2,k}\\
& \leq & c \sum_{k=m}^{l} (k+\ln n_{1,k}) (r2^k+1)^{d_1}2^{d_2m-\frac12 (r+d_1)(k-m)}                     \\
& \leq & c \log n \sum_{k=m}^{l} 2^{d_1k} 2^{d_2m -\frac12 (r+d_1)(k-m)}\\
& \leq & c \log n \; 2^{(d_1+d_2)m}\sum_{k=m}^{l} 2^{-\frac12 (r-d_1)(k-m)}
\asymp n\log n ,
\end{eqnarray*}
since $d_1k+d_2m-\frac12 (r+d_1)(k-m)=(d_1+d_2)m-\frac12 (r-d_1)(k-m)$.
This means for $r<d_1$ the cost is of order $n$
and for $r>d_1$ it is of order $n\log n$, such that a rescaling of $n$
leads to the proposed rate.
In the case $r=d_1$ the cost is $\mathcal{O}(n(\log n)^2)$,
so together with (\ref{r_gl_d1}) we get the additional log-factor in the 
convergence rate.
Now the upper bound of Theorem \ref{th:main} is proved.\\[8pt]

Note that the proof of the upper bound was carried out in terms of query 
errors and can easily be translated into an explicit
quantum algorithm for parametric integration.
The algorithm has the following form and uses the sequences of optimal
algorithms $A_{sum}(\cdot, n, N)$ and $A_{int}(\cdot,n)$ for quantum summation
and quantum integration. For a given $n$ we recall the needed parameters, 
which are $m=\left\lfloor \frac{1}{d_1+d_2}(\log_2 n)+1 \right\rfloor$, 
starting level $\widetilde{m}:=m$ if $r\geq d_1$ and zero otherwise, final level 
$l:= \lceil (1+d_2/r)m \rceil$ if $r \geq d_1$ and $(1+d_2/d_1)m\rceil -p$ 
otherwise, where $p:= \lfloor (\log_2 m)/d_1 \rfloor$. We have
$n_{1,k} = (r2^k+1)^{d_1}$,
$M_k :=\left\lceil 8(k+3) \ln 2 + 8\ln n_{1,k} \right\rceil$, and
$n_{2,k}:= \lceil 2^{d_2m-\frac12 (r+d_1)(k-m)} \rceil$, if $r \geq d_1$ and
$\lceil M_k^{-1} 2^{(d_1+d_2)m-d_1k -\frac12(d_1-r)(l-k)} \rceil$  
otherwise.
Finally we have $N_k :=  2^{rk d_2}n_{2,k}^{d_2}$. Now the algorithm
$A_{parint}(f,n)$ is the following:
\begin{enumerate}
\item \texttt{Starting level $\widetilde{m}$:}
For all $s\in \Lambda_{\widetilde{m}}^{(1)}$ do:
\begin{itemize}
  \item[(a)] If $r\geq d_1$, compute $M_{\widetilde{m}}$ times  
             $A_{\widetilde{m},s} := A_{int}(f(s,\cdot), n_{2,\widetilde{m}})$ 
             and let $\xi_{\widetilde{m}} (s)$ be the median of these 
             $M_{\widetilde{m}}$ results
  \item[(b)] If $r < d_1$, compute $M_{\widetilde{m}}$ times  
             $A_{\widetilde{m},s} := A_{sum}(\Gamma_{\widetilde{m},s} f, n_{2,\widetilde{m}},N_{\widetilde{m}})$ 
             and let $\xi_{\widetilde{m}} (s)$ be the median of these 
             $M_{\widetilde{m}}$ results
\end{itemize} 
\item \texttt{Finer levels:}
For $k=\widetilde{m}+1, \ldots , l$ do:
\begin{itemize}
 \item For all $s \in \Lambda_k^{(1)} \setminus \Lambda_{k-1}^{(1)}$ do:\\
       $M_k$ times compute 
       $A_{k,s}:= c_1 2^{-rk} A_{sum}\left( c_1^{-1}2^{rk}\Gamma_{k,s} f,n_{2,k}, N_k \right)$  
       and let $\xi_k (s)$ be the median of these $M_k$ results   
 \item For $s\in \Lambda_{k-1}^{(1)}$ do: 
        $\xi_k(s):= 0$
\end{itemize}
\item \texttt{Final approximation:}
$A_{parint}(f,n) :=P_{\widetilde{m}}^{(1)} \xi_{\widetilde{m}} +\sum_{k=\widetilde{m}+1}^l (P_k^{(1)}-P_{k-1}^{(1)}) \xi_k$
\end{enumerate}
In step $2$ the function $\Gamma_{k,s} f \in L_\infty^{N_k}$ is scaled by
$c_1^{-1}2^{rk}$ and the result then rescaled to make sure that the algorithm
$A_{sum}$ is applied to a function with $L_{\infty}^{N_k}$-norm smaller 
or equal to one.
%%%%%%%%%%%%%%%%%%%%%%%%%%%%%%%%%%%%%%%%%%%%%%%%%%%%%%%%%%%%%%%%%%%%%%%%%

\section{Lower Bound}
\label{sec:4}

In this Section we first cite a general result for lower bounds on the quantity
$e_n^q(S,F)$ and then we apply this result to the case of parametric 
integration.

Let $D$ and $K$ be nonempty sets, let $L\in \mathbb{N}$ and let to each 
\[
u= (u_0, \ldots, u_{L-1})\in \{0,1\}^L
\] 
an $f_u \in \mathcal{F}(D,K)$ be assigned
such that the following is satisfied:\\[8pt]
{\bf Condition (I):} There are functions $g_0, g_1 \in \mathcal{F}(D,K)$ and a
decomposition $D = \bigcup_{l=0}^{L-1} D_l$ with $D_l \cap D_{l'} = \emptyset
\enspace (l\neq l')$
such that for $t\in D_l$
\begin{eqnarray*}
 f_u(t) & = & \left\{ \begin{array}{ll}
                   g_0(t)  & \text{if} \enspace u_l =0 \\                    
                   g_1(t)  & \text{if} \enspace u_l =1. 
                      \end{array}
             \right.
\end{eqnarray*}
Next we define the function $\rho (L,l,l')$ for 
$L\in \mathbb{N}, 0\leq l\neq l' \leq L$ by
\[
\rho(L,l,l') = \sqrt{\frac{L}{|l-l'|}} +\frac{\min_{j=l,l'} \sqrt{j(L-j)}}{|l-l'|}.
\]

Note that $j(L-j)$ is minimized iff $|L/2-j|$ is maximized. 
For $u\in \{0,1\}^L$ we set
$|u| = \sum_{l=0}^{L-1} u_l$. Then we have the following
\begin{lemma}
\label{low}
There is a constant $c_0 > 0$ such that the following holds: Let $D,K$ be
nonempty sets, let $F\subset \mathcal{F}(D,K)$ be a set of functions, $G$ a normed
space, $S:F\rightarrow G$ a function, and $L \in \mathbb{N}$. Suppose 
$(f_u)_{u\in \{0,1\}^L} \subseteq \mathcal{F}(D,K)$ is a system of functions satisfying
condition (I). Let finally $0\leq l\neq l' \leq L$ and assume that
\[
   f_u\in F\quad \text{whenever}\quad |u|\in\{l,l'\}.
\]
Then
\[  
  e_n^q(S,F) \geq \frac12 \min \left\{\| S(f_u)-S(f_{u'})\|: |u|=l, |u'|=l' \right\}
\]
for all $n$ with
\[
   n\leq c_0 \rho(L,l,l').
\]
\end{lemma}
A proof of Lemma \ref{low} can be found in Heinrich \cite{qu:hein:2001a}.
With the help of this Lemma we can now prove the lower bound for parametric
integration:

Let $c_0$ be the constant from Lemma \ref{low}, and let $d=d_1+d_2$. For $n\in
\mathbb{N}$ we choose an even number $m\in \mathbb{N}$ such that
\begin{equation}
\label{eq:L}
L = m^d \geq \sqrt{4c_0^{-2} n^2 +4}.
\end{equation}
Let 
\[
   i = j_1m^{d-1} + j_2m^{d-2} + \ldots + j_{d-1} m + j_d
\]
for $i=0,\ldots ,L-1$. Let
\[
   D_i = \prod_{l=1}^d \left[ \frac{j_l}{m},\frac{j_l+1}{m}\right],\quad
   D = \bigcup_{i=0}^{L-1} D_i .
\]
Define the functions $\psi_i \in C^{\infty}(D)$ by
\begin{eqnarray*}
   \psi_i (s_1,\dots , s_{d_1},t_{d_1+1},\ldots ,t_{d_1+d_2}) 
  & = & \psi_i^{(1)} (s_1,\dots , s_{d_1}) 
     \cdot \psi_i^{(2)} (t_{d_1+1},\ldots ,t_{d_1+d_2}) \\ 
  & = & \prod_{l=1}^{d_1} \eta (m s_l -j_l) 
     \cdot \prod_{l=d_1+1}^{d_1+d_2} \eta (m t_l -j_l)
\end{eqnarray*}
with
\begin{eqnarray*}
 \eta (x) & = & \left\{ \begin{array}{ll}
                   e^{-\frac{1}{x(1-x)}}  &, 0 < x < 1 \\                    
                   0  & \text{, otherwise.}  
                      \end{array}
             \right.
\end{eqnarray*} 
For $k=\sum_{l=1}^{d_1+d_2} k_l$ it holds
\begin{eqnarray*}
\frac{\partial \psi_i(s,t)}{\partial s_1^{k_1}\partial s_2^{k_2}\dots \partial t_{d_1+d_2}^{k_{d_1+d_2}}} 
& = & \prod_{l=1}^{d_1} \frac{d^{k_l}}{ds_l^{k_l}} \eta (ms_l -j_l)\cdot
      \prod_{l=d_1+1}^{d_1+d_2} \frac{d^{k_l}}{dt_l^{k_l}} \eta (mt_l -j_l)\\
& = & \prod_{l=1}^{d_1} \eta^{(k_l)}(ms_l -j_l) m^{k_l}\cdot
      \prod_{l=d_1+1}^{d_1+d_2} \eta^{(k_l)}(mt_l -j_l) m^{k_l}\\
& = & m^k \prod_{l=1}^{d_1}\eta^{(k_l)}(ms_l -j_l)
      \cdot \prod_{l=d_1+1}^{d_1+d_2}\eta^{(k_l)}(mt_l -j_l) .
\end{eqnarray*}
From this we get
\begin{eqnarray*}
\| \psi_i \|_{C^r(D)} & = & \max_{\sum_{l=1}^{d_1+d_2}k_l \leq r}
\sup_{(s,t)\in D} \left| \frac{\partial^k \psi_i(s,t)}{\partial s_1^{k_1}\partial s_2^{k_2}\dots \partial t_{d_1+d_2}^{k_{d_1+d_2}}} \right| \\
& = & \max_{\sum_{l=1}^{d_1+d_2}k_l \leq r} m^k 
\prod_{l=1}^{d_1} \sup_{s_l \in [0,1]} |\eta_{(k_l)}(s_l)| 
\prod_{l=d_1+1}^{d_1+d_2} \sup_{t_l \in [0,1]} |\eta_{(k_l)}(t_l)| \\
& \leq & \gamma m^r,
\end{eqnarray*}
where 
\[
\gamma = \max_{\sum_{l=1}^{d_1+d_2}k_l \leq r}
\prod_{l=1}^{d_1}\sup_{s_l \in [0,1]} |\eta_{(k_l)}(s_l)| \cdot
\prod_{l=d_1+1}^{d_1+d_2}\sup_{t_l \in [0,1]} |\eta_{(k_l)}(t_l)| .
\]
Therefore, setting
$\hat{\psi_i} := \frac{1}{\gamma m^r} \psi_i$, we have 
$f_u := \sum_{i=0}^{L-1} u_i \hat{\psi_i} \in F$
for all $u_i \in \{0,1\}, i=0,\ldots ,L-1$. Since the $\psi_i$ have disjoint
support, $f_u$ satisfies condition (I). Let 
$\sigma_0 = \int_0^1 \eta(x)dx > 0$. It is easy to show that
\[ 
  \int_D \psi_i(s_1,\ldots ,t_{d_1+d_2})\; ds_1\ldots dt_{d_1+d_2} 
   = \frac{\sigma_0^d}{m^d}.
\] 
Let now $l:=L/2-1$ and $l':=l+1=L/2$. Then with
(\ref{eq:L}) it follows that
\[
   \rho (L,l,l') > \min_{j=\{ l,l'\} } \sqrt{j(L-j)} = \sqrt{\left( \frac{L}{2} -1
\right) \left( \frac{L}{2} +1 \right)} = \sqrt{\frac{L^2}{4}-1} \geq c_0^{-1} n.
\]
Since
\begin{eqnarray*}
\lefteqn{\min \left\{ \| S(f_u) - S(f_{u'}) \|_{C(D_1)} \; | \; |u| = l, |u'|=l' \right\}} \\  
  & = & \left\| S(\hat{\psi_0}) \right\|_{C(D_1)} =
   \frac{1}{\gamma m^r} \frac{\sigma_0^{d_2}}{m^{d_2}} 
   \left\| \psi_0^{(2)} (t_{d_1+1},\ldots ,t_{d_1+d_2})\right\|_{C(D_1)} \\ 
 & = & e^{-4d_1}\sigma_0^{d_2} \gamma^{-1} m^{-(r+d_2)}
   = C \left( m^{d_1+d_2} \right)^{-\frac{r+d_2}{d_1+d_2}}\\
 & = & C' n^{-\left( \frac{r}{d_1+d_2} +\frac{d_2}{d_1+d_2} \right)} ,
\end{eqnarray*}
from Lemma \ref{low} we get a general lower bound for parametric integration 
and the lower bound of Theorem \ref{th:main} for $r\geq d_1$.

We get a lower bound of $n^{-r/d_1}$ for approximation by simply 
choosing $d_2 = 0$, and since parametric integration can never have a better 
convergence rate than approximation, this gives us the lower bound of 
Theorem \ref{th:main} for the case $r < d_1$.

\section{Comments}
\label{sec:5}

We have solved the problem of the quantum complexity of parametric integration
for the class $C^r(D)$ by providing upper and lower bounds, where the rates 
match up to a logarithmic factor.
Now we compare our results to the known results for deterministic and
Monte Carlo methods stated in Section \ref{sec:2.3}. We again assume $r\geq 1$.
Then the optimal quantum
rates are always better than the optimal deterministic rate, except for
the case $d_2=0$, where we have equal rates, and the
following table provides a comparison of the deterministic, Monte Carlo and 
quantum case (without log- factors).
\[
\begin{tabular}{|l |l |l |l |} \hline 
  & $e_n^{det}(S,F)$ & $e_n^{mc}(S,F)$  & $e_n^q(S,F)$ \\ \hline 
  $r\geq d_1$ & $n^{-r/(d_1+d_2)}$ & $n^{-(r+d_2/2)/(d_1+d_2)}$ &  $n^{-(r+d_2)/(d_1+d_2)}$ \\ \hline 
  $d_1/2 \leq r < d_1$ & $n^{-r/(d_1+d_2)}$ & $n^{-(r+d_2/2)/(d_1+d_2)}$ &  $n^{-r/d_1}$ \\  \hline 
  $r < d_1/2$ & $n^{-r/(d_1+d_2)}$ & $n^{-r/d_1}$ & $n^{-r/d_1}$  \\ \hline
\end{tabular}
\]
We have to distinguish three different situations, depending on the relation
of the problem parameters $r$ and $d_1$. For $r<d_1/2$ the quantum rate
provides an improvement over the deterministic rate, but it is as fast as 
the Monte Carlo rate. For this parameter constellation
both the Monte Carlo and the quantum algorithm achieve the optimal rate of 
approximation which is the same in both cases.

When we have $d_1/2 \leq r < d_1$, then the quantum rate is still the
optimal rate of approximation, but the Monte Carlo rate is slower, which
leads to the superiority of the quantum rate for this situation.

For the case $r\geq d_1$ we still have a better performance of the
quantum algorithm as compared to Monte Carlo.

Summarizing the discussion we can say that the quantum rate is always better
than the deterministic rate, it is always at least as good as the Monte
Carlo rate, and for $r\geq d_1/2$ it is better than the optimal rate
of Monte Carlo algorithms.  

Let us finally consider our problem in the bit model. Referring to 
\cite{qu:hein:2001a} and \cite{qu:hein:2002}, we find that for our algorithm
the number of qubits needed is $\mathcal{O}(\log n)$, the number of quantum 
gates is $\mathcal{O}(n\log n)$ and the number of measurements is 
$\mathcal{O}((\log n)^3 \log \log n)$.

\section*{Acknowledgement}

I would like to thank Stefan Heinrich for scientific guidance, discussions,
and suggestions concerning this paper. 

%%%%%%%%%%%%%%%%%%%%%%%%%%%%%%%%%%%%%%%%%%%%%%%%%%%%%%%%%%%%%%%%%%%%%%%%%%
\bibliographystyle{plain}

\end{document}